\documentclass[a4paper,UKenglish,cleveref, autoref, thm-restate]{lipics-v2021}
\usepackage{graphicx}
\usepackage{hyperref}
\usepackage{subfiles}
\usepackage{amsmath}
\usepackage{cleveref}
\usepackage{dsfont}
\usepackage{ifsym}
\usepackage{comment}
\usepackage[dvipsnames]{xcolor}
\usepackage{boxedminipage}
\usepackage[ruled,vlined]{algorithm2e}

\RestyleAlgo{ruled}

\SetKwComment{Comment}{/* }{ */}

\title{Exact Matching: Algorithms and Related
Problems} 

\author{Nicolas {El Maalouly}}{Department of Computer Science, ETH Z\"{u}rich, Switzerland }{nicolas.elmaalouly@inf.ethz.ch}{0000-0002-1037-0203}{}

\authorrunning{N. El Maalouly} 

\Copyright{Nicolas El Maalouly} 



\relatedversiondetails{Full Version}{https://arxiv.org/abs/2203.13899} 

\acknowledgements{I want to thank Ola Svensson for introducing me to the Exact Matching problem, for defining the Top-k Perfect Matching problem and for helpful discussions and feedback.}

\nolinenumbers 

\EventEditors{Petra Berenbrink, Mamadou Moustapha Kant\'{e}, Patricia Bouyer, and Anuj Dawar}
\EventNoEds{4}
\EventLongTitle{40th International Symposium on Theoretical Aspects of Computer Science (STACS 2023)}
\EventShortTitle{STACS 2023}
\EventAcronym{STACS}
\EventYear{2023}
\EventDate{March 7--9, 2023}
\EventLocation{Hamburg, Germany}
\EventLogo{}
\SeriesVolume{254}
\ArticleNo{36}

\begin{document}

\maketitle

\begin{abstract}
In 1982, Papadimitriou and Yannakakis introduced the \emph{Exact Matching} (EM) problem where given an edge colored graph, with colors red and blue, and an integer $k$, the goal is to decide whether or not the graph contains a perfect matching with exactly $k$ red edges. Although they conjectured it to be \textbf{NP}-complete, soon after it was shown to be solvable in randomized polynomial time in the seminal work of Mulmuley et al., placing it in the complexity class \textbf{RP}. Since then, all attempts at finding a deterministic algorithm for EM have failed, thus leaving it as one of the few natural combinatorial problems in \textbf{RP} but not known to be contained in \textbf{P}, and making it an interesting instance for testing the hypothesis $\textbf{RP}=\textbf{P}$. 
Progress has been lacking even on very restrictive classes of graphs despite the problem being quite well known as evidenced by the number of works citing it. 

In this paper we aim to gain more insight into EM by studying a new optimization problem we call \emph{Top-k Perfect Matching} (TkPM) which we show to be polynomially equivalent to EM. By virtue of being an optimization problem, it is more natural to approximate TkPM so we provide approximation algorithms for it. Some of the approximation algorithms rely on a relaxation of EM on bipartite graphs where the output is required to be a perfect matching with a number of red edges differing from $k$ by at most $k/2$, which is of independent interest and generalizes to the \emph{Exact Weight Perfect Matching} (EWPM) problem. We also consider parameterized algorithms and show that TkPM can be solved in FPT time parameterized by $k$ and the independence number of the graph. This result again relies on new tools developed for EM which are also of independent interest.

\end{abstract}

\newpage
\section{Introduction}


Deciding whether randomization adds power to sequential algorithms is an central problem in complexity theory. The main question there is whether $\textbf{P} = \textbf{RP}$ which remains a big open problem and is tied to other important questions in the field \cite{kabanets2004derandomizing}. Only few natural problems are known to be in \textbf{RP} while no deterministic algorithms are known for them.
Exact Matching (EM), defined in 1982 by Papadimitriou and Yannakakis~\cite{papadimitriou1982complexity}, is one such problem. 

\vspace{5pt}

\noindent\vspace{5pt}\begin{boxedminipage}{\textwidth}
\textsc{Exact Matching (EM)}

\textbf{Input:} A graph $G$, with each edges colored red or blue, and integer $k$.

\textbf{Task:} Decide whether there exists a perfect matching $M$ in $G$ with exactly $k$ red edges.

\end{boxedminipage}
At the time of its introduction it was conjectured to be \textbf{NP}-complete. Only a few years later, however, it was shown to be in \textbf{RP} by Mulmuley, Vazirani and Vazirani~\cite{mulmuley1987matching}, which makes it unlikely to be \textbf{NP}-hard. In fact, in was even shown to be in \textbf{RNC} which is defined as the class of decision problems allowing an algorithm running in polylogarithmic time\footnote{in the following, $n$ denotes the number of vertices of the input graph} (i.e., $O(\log n^c)$ for some constant $c>0$) using polynomially many parallel processors, while having additional access to randomness (we refer the interested reader to~\cite{complexitybook} Chapter~12 for a formal definition).  
Derandomizing matching problems from this complexity class is also a big open problem~\cite{svensson2017matching}. This makes EM even more interesting since randomness allows it to be efficiently parallelizable while it even remains difficult to solve sequentially without such access to randomness.

The interest in EM is evidenced by the numerous works that cite it.
These include works on the parallel computation complexity of the matching problem~\cite{svensson2017matching}, planarizing gadgets for perfect matchings~\cite{gurjar2012planarizing}, multicriteria optimization~\cite{grandoni2010optimization}, matroid intersection~\cite{camerini1992random}, DNA sequencing~\cite{blazewicz2007polynomial}, binary linear equation systems with small hamming weight~\cite{arvind2016solving}, recoverable robust assignment~\cite{fischer2020investigation} in addition to generalizations of the problem with multiple color constraints~\cite{berger2011budgeted,mastrolilli2012constrained,mastrolilli2014bi,stamoulis2014approximation}. 
Despite that, deciding whether EM is in \textbf{P} has remained an open problem for almost four decades and little progress has been made even for very restricted classes of graphs, thus highlighting the surprising difficulty of the problem.

\subsection{Prior Work.}  
\subparagraph{Restricted Graph Classes.}  
When it comes to restricted classes of graphs, results go in two directions. The first is the sparse graphs regime where in the extreme case we have trees for which EM can be solved by a simple dynamic program (DP). This can also be generalized to bounded tree-width graphs. Such a DP has not been explicitly given in the literature but would be easy to construct by keeping track of how every edge in a bag is matched (not yet matched, matched outside the bag or matched inside the bag) as well as the total number of red edges in the matching so far. It is easy to see that the number of possible states is at most $O(3^{tw} \cdot n)$ (where $n$ comes from the possible number of red edges in the matching) resulting in an FPT algorithm parameterized by the tree width of the graph. 
Continuing with sparse graph classes, EM is also known to be solvable for planar graphs~\cite{yuster2012almost} by relying on the existence of Pfaffian orientations to derandomize the \textbf{RNC} algorithm. The same techniques used for this derandomization also allow for computing the matching generating function (see \cite{godsil2017algebraic} Chapter 1 for a definition) and can be generalized to other graph classes such as $K_{3,3}$-minor free graphs and graphs embeddable on a surface of bounded genus~\cite{genus}. Computing the matching generating function was recently shown to be \#\textbf{P}-hard already for $K_8$-minor free graphs \cite{curticapean2021parameterizing} so these results do not generalize much further and are restricted to very sparse graphs.

The second direction is dense graph classes. Here it is known that EM is in \textbf{P} for complete and complete bipartite graphs, i.e., graphs of independence number $\alpha =1$ and bipartite graphs of bipartite independence number\footnote{ The \emph{bipartite independence number} of a bipartite graph $G$ equipped with a bipartition of its vertices is defined as the largest number $\beta$ such that $G$ contains a \emph{balanced independent set} of size $2\beta$, i.e., an independent set using exactly $\beta$ vertices from each side of the bipartition.} $\beta =1$. In fact, these results are already non-trivial and at least four different articles have appeared on resolving them~\cite{karzanov1987maximum,yi2002matchings,geerdes,gurjar2017exact}. Very recently, however, El Maalouly and Steiner \cite{elmaalouly2022exact} pushed the boundary of positive results further by showing that EM is in \textbf{P} for all graphs of bounded independence number and all bipartite graphs of bounded bipartite independence number.

\subparagraph{Generalizations.}  
As mentioned above, prior work also considered a generalization of the problem to multiple color constraints, known as Bounded Color Matching (BCM).

\vspace{5pt}
\noindent\vspace{5pt}\begin{boxedminipage}{\textwidth}
\textsc{Bounded Color Matching (BCM)}

\textbf{Input:} A weighted and edge-colored (with colors $c_1,...,c_l$) graph $G$ and integers $k_1,...,k_l$.

\textbf{Task:} Find a maximum weight matching $M$ in $G$ with at most $k_i$ edges of color $i$ for all $i\in \{1,...,l\}$.

\end{boxedminipage}
BCM is known to be NP-hard \cite{rusu2008maximum}. Mastrolilli and Stamoulis \cite{mastrolilli2014bi}  provide bi-criteria approximation schemes which give an approximately maximum matching with small constraint violations. Stamoulis \cite{stamoulis2014approximation} also gives a $1/2$-approximation for the objective with no constraint violations. No prior work considered bounds on the constraint violations while requiring an optimal objective, i.e., a perfect matching if the graph is unweighted.
For EM, however, Yuster~\cite{yuster2012almost} proved that given an instance of the problem, one can decide in polynomial time that either $G$ contains no perfect matching with exactly $k$ red edges, or one can compute an \emph{almost} perfect matching (i.e., of size at least $\lfloor \frac{n}{2}\rfloor-1$) containing $k$ red edges.
This means that the techniques used in the bi-criteria approximation of BCM do not provide much further insight into solving EM since they relax the perfect matching requirement. 

Another way to generalize the problem is to have a weighted instead of edge-colored graph, and require the output perfect matching to have an exact weight.

\vspace{5pt}

\noindent\vspace{5pt}\begin{boxedminipage}{\textwidth}
\textsc{Exact Weight Perfect Matching (EWPM)}

\textbf{Input:} A weighted graph $G$ and integer $W$.

\textbf{Task:} Find a perfect matching $M$ in $G$ with $w(M) = W$.

\end{boxedminipage}
EWPM can be reduced to EM if the edge weights are polynomial in the input size but is known to be NP-hard for exponential weights~\cite{gurjar2012planarizing}.
This makes approximation algorithms that aim to minimize the constraint violation even more desirable for EWPM.


\subsection{Our contribution.}

\subparagraph{Exact Matching.}
We provide an algorithm for a relaxed version of EM on bipartite graphs where we require the output to be a perfect matching and allow for a constraint violation that is a constant fraction of $k$, $0.5$ in this case. 
\begin{theorem}\label{th:EMapprox}
There exists a deterministic polynomial time algorithm that, given a "Yes" instance of EM on a bipartite graph, outputs a perfect matching $M$ with $0.5k \leq |R(M)| \leq 1.5k$, where $|R(M)|$ is the number of red edges in $M$.
\end{theorem}
This can also be seen as an attempt to approximate the EM problem without relaxing the perfect matching constraint.
This type of approximation is the first of its kind for EM. Note that in light of the above mentioned result by Yuster~\cite{yuster2012almost} (i.e., an algorithm that outputs an \emph{almost} perfect matching containing $k$ red edges) one would think that it should not be too difficult to find an algorithm for the relaxed version of EM with a constraint violation of only one red edge. However, the perfect matching requirement seems to be intrinsic to the difficulty of the problem (given the simplicity of Yuster's algorithm) and many attempts at improving the constraint violation of \Cref{th:EMapprox} have failed so far.
We also show that the approximation algorithm works for the more general problem of EWPM, only loosing $1/poly(n)$ in the approximation factor for exponential weights.
\begin{corollary}\label{cor:EWPMapprox}
There exists a deterministic polynomial time algorithm that, given a "Yes" instance of EWPM on a bipartite graph with input weights bounded by a polynomial (resp. exponential) function of the input size, outputs a perfect matching $M$ with $0.5W \leq w(M) \leq 1.5W$ (resp. $(0.5-1/poly(n))W \leq w(M) \leq (1.5+1/poly(n))W$).
\end{corollary}
We also introduce a new way of finding alternating cycles with certain color and weight properties in FPT time parameterized by the number of edges in the cycles. 
\begin{proposition}\label{prop:smallsetcycles}
Let $G = (V,E,w)$ be an edge colored and weighted graph with edge colors red and blue, and let $M$ and $M'$ be two perfect matchings in $G$ and $\mathcal{C} = M \Delta M'$ s.t. $|E(\mathcal{C})| \leq L$ for some integer $L$. Then there exists an algorithm running in time $f(L)poly(n)$ (for $f(L) = L^{O(L)}$) that, given $G$ and $M$ as input, outputs a perfect matching $M''$ in $G$ with $w(M'') \geq w(M')$ and $|R(M'')| = |R(M')|$.
\end{proposition}
This allows us to get an FPT algorithm for EM, parameterized by the circumference of the graph. 
\begin{theorem}\label{th:fptcirc}
There exists a deterministic FPT algorithm, parameterized by the circumference\footnote{The circumference of a graph is the length of any longest cycle in the graph.} of the graph, for the Exact Matching problem in general graphs.
\end{theorem}

\subparagraph{Top-k Perfect Matching.}
The above studied problems suffer from the fact that they are not optimization problems (due to the exactness constraint which requires the optimization of more than one objective) and are thus less natural to approximate.
For this reason, we study a new matching problem called Top-$k$ Perfect Matching.

\vspace{5pt}
\noindent\vspace{5pt}\begin{boxedminipage}{\textwidth}

\textsc{Top-$k$ Perfect Matching (TkPM)}

\textbf{Input:} A weighted graph $G$ and integer $k$.

\textbf{Task:} Find a perfect matching in $G$ maximizing the top-$k$ weight function.

\end{boxedminipage}
Here the top-k weight function is defined as the sum of the weights of the $k$ highest weight edges in the matching.
To our knowledge, this problem has not yet been considered in the literature, but similar types of optimization objectives have been used for other problems such as $k$-clustering and load balancing \cite{chakrabarty2019approximation}.
We show that this problem can also be reduced to EM (in deterministic polynomial time) when the edge weights are polynomially bounded in the input size.
\begin{theorem}\label{th:TkPMtoEM}
$TkPM \leq_p EM$ for polynomially bounded weights.\footnote{For two problems $A$ and $B$, $A \leq_p B$ means that $A$ is reducible to $B$ in deterministic polynomial time and $A \equiv_p B$ implies both $A \leq_p B$ and $B \leq_p A$. 
} 
\end{theorem}
This puts TkPM with polynomial weights in the class \textbf{RP} and it remains open whether or not it is in \textbf{P}, thus making it another natural problem in this category. Interestingly, a recent result shows that EM can in turn be reduced to TkPM, making the two problems polynomially equivalent.
\begin{lemma}[from \cite{elmaalouly2022exacttopkequiv}]\label{th:EMtoTkPM}
$EM \leq_p TkPM$ for polynomially bounded weights.
\end{lemma}
This means that progress on TkPM not only provides further insight into EM, but could also help solve it directly.
As previously mentioned, the main advantage of TkPM over the other studied variants of EM is that it is an optimization problem, i.e., we are maximizing a single objective function. This makes it more suitable for approximation and we provide approximation algorithms for it.
\begin{theorem}\label{th:05approx}
There exists a deterministic polynomial time $0.5$-approximation algorithm for TkPM.
\end{theorem}
\begin{theorem}\label{th:08approx}
There exists a deterministic polynomial time $(0.8-1/poly(n))$-approximation algorithm for TkPM on bipartite graphs. 
\end{theorem}
It is interesting to note that the main tool used for the proof of \Cref{th:EMapprox} (i.e., \Cref{prop:constrainedCycle}) was originally developed to prove \Cref{th:08approx}. This shows how the study of TkPM can indeed provide insight into the EM problem.

Finally, the techniques we developed for FPT algorithms for EM so far only resulted in an FPT algorithm parameterized by the circumference of the graph. The circumference, however, is usually quite large and not very good as a parameter, so to better illustrate the use of these techniques, we combine them with techniques from \cite{elmaalouly2022exact} to show the existence of an FPT algorithm for TkPM parameterized by $k$ and $\alpha$ (the independence number of the input graph), and an FPT algorithm for TkPM on bipartite graphs parameterized by $k$ and $\beta$ (the bipartite independence number of the input graph).
\begin{theorem}\label{th:FPTkalpha}
There exists a deterministic algorithm for TkPM running in time $f(k,\alpha)poly(n)$ where $f(k,\alpha) = (k4^{\alpha})^{O(k4^{\alpha})}$ and $\alpha$ is the independence number of the input graph.
\end{theorem}
\begin{theorem}\label{th:FPTkbeta}
There exists a deterministic algorithm for TkPM on bipartite graphs running in time $f(k,\beta)poly(n)$ where $f(k,\beta) = (k\beta)^{O(k\beta)}$ and $\beta$ is the bipartite independence number of the input graph.
\end{theorem}

\subsection{Organization of the paper.}

The remainder of this paper is organized as follows: 
In \Cref{sec:Prel} we present the basic definitions and conventions we use throughout the paper.
In \Cref{sec:EM} and \Cref{sec:TkPM} we study EM and TkPM respectively, both from the perspectives of approximation and parameterized algorithms.
Finally in \Cref{sec:conc} we conclude the paper and provide some open problems.

\section{Preliminaries}\label{sec:Prel}

Due to space restrictions, proofs of statements marked ($\star$) have been deferred to
the appendix. All graphs considered are simple.
For a red/blue edge colored graph $G$ and $G'$ a subgraph of $G$, we define $R(G')$ (resp. $B(G')$) to be the set of red (resp. blue) edges in $G'$ and $w(G')$ to be the sum of the weights of edges in $G'$.
Undirected cycles are considered to have an arbitrary orientation. For a cycle $C$ and $u,v \in C$, $C[u,v]$ is defined as the path from $u$ to $v$ along $C$ (in the fixed but arbitrarily chosen orientation if $C$ is undirected). 
Given a matching $M$, $C$ is called $M$-alternating if for any two adjacent edges in $C$, one of them is in $M$ and the other is not. An $e$ edge is called a matching edge if $e \in M$ and a non-matching edge if $e \notin M$. 

We always assume that for problems on weighted graphs, the input weights are given as positive integers and their encoding size is part of the input (i.e., they can be at most exponential in the input size if they are encoded in binary). We use $w$ to refer to the set of weights in a weighted graph $G= (V,E,w)$. We always consider a strict ordering on the edges in which the edges are ordered by decreasing weight with ties broken arbitrarily (but the ordering is fixed for a given graph and weight function).
The top-k weight function $w^k(E)$ for a set of edges $E$ is defined as the sum of the first $k$ edges from $E$ in the edge ordering of the graph, i.e., $w^k(E) = \sum_{i \in \{1...k\}}w(E(i))$ where $E(i)$ is the $i$-th edge from $E$ in the edge ordering of the graph.

\section{Exact Matching} \label{sec:EM}

\subsection{Approximation Algorithms}

In this section, we aim to prove \Cref{th:EMapprox} by developing a deterministic polynomial time algorithm for EM where we require the output to be a perfect matching (abbreviated PM) and allow for a constraint violation that is a constant fraction of $k$. More precisely we require the output PM to have between $0.5k$ and $1.5k$ red edges. The main tool we use is the following proposition which allows us to increase the number of red edges of a PM without adding too many such edges.

\begin{proposition} \label{prop:constrainedCycle}
Let $G := (V, E)$ be an edge weighted directed graph containing a directed cycle $C$ with 
$w(C) > 0$ and $C$ contains at most $k$ edges having strictly positive weight. 
There exists a deterministic polynomial time algorithm that, given $G$, finds a directed cycle in $G$ with the same properties as $C$. 
\end{proposition}

\begin{proof}[Proof of \Cref{prop:constrainedCycle}]\label{prop:constrainedCycle:proof}
For simplicity, we will flip the sign of all weights so that we are looking for a negative cycle which can be found by a shortest path algorithm. In the following we will use the Bellman-Ford algorithm which relies on a dynamic program (DP) to compute the distance between any two nodes in the graph \cite{bellman1958routing}.
By adding an extra constraint variable to the DP, we are also able to compute the shortest path weights for paths that fulfill some bound on the constraint. More formally we start with the normal update rule for the Bellman-Ford algorithm:

$$ d(s, v) = \min_{u \in V}\{d(s, u) + \Bar{w}(u,v)|(u, v) \in E\}$$
where $\Bar{w}(u,v) = -w(e)$ (i.e., we flip the sign of the weights) for $e=(u,v)$ and $d(s,v)$ is the distance from $s$ to $v$ where the length of an edge is given by its weight $\Bar{w}$ (note that every vertex is considered to have a self loop of weight $0$).
We modify it to include the constraint variable (with an extra dimension in the table entries of the DP to account for it):

$$ d(s, v, c) = \min_{u \in V}\{d(s, u, c - \mathds{1}_{\Bar{w}(u,v) < 0}) + \Bar{w}(u,v)|(u, v) \in E\}$$
where $\mathds{1}$ is the indicator variable which takes value $1$ if the condition is true and $0$ otherwise, so the constraint variable is decreased every time the path uses a red edge.
The entries $d(s, v, c)$ are initialized to $\infty$ for all $s, v \in V(G)$ and $c \in \{-1,0,1,2,...,k\}$, except for the entries of the form $d(s,s,c)$, for all $s \in V(G)$ and $c \in \{0,1,2,...,k\}$, which are initialized to $0$. This way, after running the update rule on the DP until convergence or until some entry of the form $d(s,s,c)$ becomes negative (i.e., a strictly negative cycle is detected), the value of $d(s, v, c)$ corresponds to the weight of the shortest path from $s$ to $v$, containing at most $c$ red edges, if such a path exists and is $\infty$ otherwise, unless there is a negative cycle. 
Note that the table entries can be computed iteratively, starting with entries of the form $d(s,v,0)$ (the computation is the same as the regular Bellman-ford algorithm but with the new update rule) then increasing $c$ by $1$ every time.
Finally observe that if a strictly negative cycle $C$ containing at most $k$ edges of strictly negative weight (i.e., positive in the original edge weight before the sign flip) exists, at least one of the entries of the form $d(s,s,c)$ for $s \in C$ and $0\leq c \leq k$ will become negative (since the shortest path from $s$ to itself should have negative length). Such a cycle is guaranteed by the conditions of the proposition and computing it can be done by a standard modification of the DP that keeps track of the last used edge for each updated entry. The output cycle is guaranteed to be strictly positive and have at most $k$ strictly positive weight edges (in the original graph before the sign flip).
Note that the running time of the DP is polynomial in the number of table entries, which in turn is polynomial in the size of the input graph.
\end{proof}

By repeatedly applying \Cref{prop:constrainedCycle} we are able to find a PM fulfilling the requirements of \Cref{th:EMapprox}.

\begin{proof}[Proof of \Cref{th:EMapprox}]
Let $M$ be a PM containing a minimum number of red edges (should be at most $k$ since we have a "Yes" instance). Note that $M$ can be computed in polynomial time by simply using a maximum weight perfect matching algorithm \cite{edmonds1965maximum}, with $-1$ weights assigned to red edges and $0$ weights assigned to blue edges. If $|R(M)| \geq 0.5k$ we are done, so suppose $|R(M)| < 0.5k$.
We define the directed graph $G'$ in the following way. We start with the bipartite input graph $G = (A \cup B, E)$ and orient the edges as follows: edges in $M$ are oriented from $A$ to $B$ and edges not in $M$ are oriented from $B$ to $A$. This way we are guaranteed that any directed cycle in the resulting graph is an $M$-alternating cycle. We also define edge weights as follows: blue edges get weight $0$, red edges in $M$ get weight $-1$ and red edges not in $M$ get weight $+1$. This way we have that for any $M$-alternating cycle $C$, $M' := M\Delta C$ is a perfect matching with $|R(M')| = |R(M)| + w(C)$. Note that $M \Delta M'$ is a set of disjoint cycles that are both $M$-alternating and $M'$-alternating.

Let $M^*$ be a solution to the EM instance, i.e., $|R(M^*)|  = k$ (which must exist since we are given a "Yes" instance). Observe that $w(M \Delta M^*) = |R(M^*)|  - |R(M)|  > 0$ so there must be a cycle $C \in M \Delta M^*$ s.t. $w(C) > 0$. Also note that $M^*$ contains exactly $k$ red edges so $M \Delta M^*$ contains at most $k$ red edges not in $M$ (i.e., edges of strictly positive weight). Finally note that the cycle $C$ is a directed cycle (since it is alternating). 
So we can use \Cref{prop:constrainedCycle} on the resulting graph to find a cycle $C'$ with $w(C')>0$ containing at most $k$ edges of strictly positive weight. Note that $w(C') \leq k$ since edges have weight at most $+1$. Now we let $M' := M \Delta C'$ (this is possible since $C'$ being a directed cycle implies that it must be an $M$-alternating cycle). Note that
$$|R(M)| < |R(M)|  + w(C') \leq |R(M)|  + k < 1.5k$$ and $|R(M')| = |R(M \Delta C')| = |R(M)|  + w(C')$.
So if $|R(M')| \geq 0.5k$ the algorithm stops and outputs $M \Delta C'$, otherwise we repeat the above procedure, with $M'$ replacing $M$, until $|R(M')| \geq 0.5k$. The running time is polynomial since the above procedure runs in polynomial time (by \Cref{prop:constrainedCycle}) and it is repeated at most $k$ times.
\end{proof}

The algorithm also works for EWPM if the weights are polynomial, using the reduction to EM, so it remains to show how to deal with exponential weights.

\begin{proof}[Proof of \Cref{cor:EWPMapprox}]\label{cor:EWPMapprox:proof}
For exponential size weights, we first scale and round them to make them bounded by a polynomial function of the input size. 
To do so, we start by deleting all edges of weight $> W$ (this is safe to do since none of these edges can be part of an optimal solution). Note that all weight encodings now have at most $\log_2(W)$ non-zero bits. Let $f(n) = 2n\cdot poly(n)$ for any desired polynomial. We re-encode the weights of all edges by only considering their $(\log_2(W)-\log_2(f(n)))$-th to $(\log_2(W))$-th bits (counting from the least significant bit) and dropping all others. We call these weights $w'$. So all weights are now encoded with at most $\log_2(f(n)) + 1$ bits, i.e., are bounded by a polynomial function of the input size. Observe that for any edge $e$, $|w(e) - w'(e)\cdot\frac{W}{f(n)}| \leq \frac{W}{f(n)}$ (the rounding error). Let $M^*$ be a solution to the EWPM instance. 
We have $$|w(M^*)-w'(M^*)\cdot\frac{W}{f(n)}| \leq \frac{nW}{2f(n)}$$ since a perfect matching contains $n/2$ edges, which implies 
$$w(M^*)\frac{f(n)}{W} - n/2 \leq w'(M^*) \leq w(M^*)\frac{f(n)}{W} + n/2$$
and using $W= w(M^*)$ we get 
$$f(n) - n/2 \leq w'(M^*) \leq f(n)+ n/2.$$
Let $M'$ be an optimal solution for EWPM with any of the following parameters: $\{f(n)-n/2,f(n)-n/2 + 1, ...,f(n)+n/2\}$ on the graph with weights $w'$ (i.e., we run an algorithm for EWPM on all parameters and output any of them if one exists). Note that $M^*$ is one possible optimal solution in this case, so a solution must exist (since we have a "Yes" instance).
Now we get $$|w(M')-W| = |w'(M')\cdot\frac{W}{f(n)}-W| \leq |w'(M')\cdot\frac{W}{f(n)}-w'(M^*)\cdot\frac{W}{f(n)}| + \frac{nW}{2f(n)} \leq 2\frac{nW}{f(n)} $$ which implies $((1-2n/f(n))W\leq w(M')\leq (1+2n/f(n))W$.

We still need an approximation for EWPM with polynomially bounded input weights. For this we use the reduction from \cite{gurjar2012planarizing} (see also \Cref{th:reductions}) to get an instance of EM where a PM with $W$ red edges corresponds to a PM of weight $W$ in the original graph. Since the weights are polynomial in the size of the input, the size of the graph remains polynomial. Now using the algorithm of \Cref{th:EMapprox} with $k:=W$ we get a PM $M$ with $0.5W \leq w(M) \leq 1.5W$ as desired.
\end{proof}




\subsection{FPT Algorithms}

In this section we start by proving \Cref{prop:smallsetcycles} which provides a new tool for finding alternating cycles with color and weight constraints in FPT time parameterized by the size of the cycles.

\begin{proof}[Proof of \Cref{prop:smallsetcycles}]
Our goal is to find a set of $M$-alternating disjoint cycles $\mathcal{C}'$ in $G$ with the same number of matching (i.e., edges in $M$) and non-matching (i.e., edges not in $M$) red edges as $\mathcal{C}$ and weights that are at least as big, i.e., for every $C \in \mathcal{C}$ there must be a $C' \in \mathcal{C}'$ such that $C'$ has the same number of matching and non-matching red edges as $C$ and $w(C') \geq w(C)$ (and vice versa, i.e., there is a one to one correspondence between the cycles in $\mathcal{C}$ and the cycles in $\mathcal{C}'$). This way we construct $M'' := M \Delta \mathcal{C'}$ s.t. $w(M'') = w(M \Delta \mathcal{C}') \geq w(M \Delta \mathcal{C}) =  w(M')$ and $|R(M'')| = |R(M \Delta \mathcal{C}')|= |R(M \Delta \mathcal{C})| = |R(M')|$.

\proofsubparagraph*{Color Coding.} 
The main tool for finding such a set of cycles is color coding \cite{alon1995color}. The idea is to color all vertices at random with $L$ colors. The probability that all vertices of $\mathcal{C}$ get different colors is only a function of $L$. This can also be achieved deterministically using a perfect hash family of size bounded by $L^{O(L)}poly(n)$, which can be guaranteed to contain at least one coloring for which all vertices of $\mathcal{C}$ have different colors (see \cite{cygan2015parameterized} Chapter 5 for more details on derandomizing color coding).

\proofsubparagraph*{Separating the Cycles.} 
Observe that for every cycle in $C \in \mathcal{C}$, the following can also be achieved in $L^{O(L)}poly(n)$ time.
\begin{itemize}
    \item Guess the set of colors $colors(C)$ of its vertices and their exact order.
    \item Guess its number of matching (i.e., in $M$) and non-matching (i.e., not in $M$) red edges.
\end{itemize}
Let $G_C$ be the graph induced on the vertices of $G$ with a color from the set $colors(C)$. Observe that $C$ is contained in $G_C$ and that the subgraphs $G_C$ for $C \in \mathcal{C}$ are all disjoint. So we can look for each cycle separately.

\proofsubparagraph*{Orienting the Cycles.} 
Since we know the colors of the vertices of $\mathcal{C}$, we can define a bipartition $(A,B)$ of $G$ by splitting the set of colors into two equal parts and letting $A$ (resp. $B$) be the vertices having a color from the first (resp. second), s.t. the cycles in $\mathcal{C}$ are alternating with respect to the bipartition (note that this is indeed possible since the cycles in $\mathcal{C}$ are $M$-alternating so they have even length). By deleting all edges with endpoints in the same part, we get a bipartite graph which contains $\mathcal{C}$. Now we can define the following orientation for the edges: edges in $M$ are oriented from $A$ to $B$ and edges not in $M$ are oriented from $B$ to $A$. This way we are guaranteed that any directed cycle in the resulting graph is an alternating cycle.

\proofsubparagraph*{From Cycles to Colorful Paths.} 
For this part and the next, we look into one cycle $C \in \mathcal{C}$ and its corresponding subgraph $G_C$. Let $(c_1,c_2,...) := colors(C)$. We first guess the edge of $C$ with start vertex from color class $c_{|colors(C)|}$ and end vertex from color class $c_1$ (this can be done in polynomial time by trying all possibilities). Then we delete all edges from $G_C$ except for the edges going from a vertex of color $c_i$ to a vertex of color $c_{i+1}$ for $i \in \{1,2,..., |colors(C)|-1 \}$. Observe that $G_C$ is now acyclic and the remaining edges of $C$ form a directed path from $s$ to $t$ in $G_C$.

\proofsubparagraph*{Finding the Paths.}
For simplicity, we will flip the sign of all weights so that we are looking for paths of minimum weight which can be found by a shortest path algorithm. Similarly to the proof of \Cref{th:EMapprox} we use a modified Bellman-Ford algorithm, so we will focus on the main difference, i.e., the update rule. 
By adding extra constraint variables to the DP, we are also able to compute the shortest path weights for paths that fulfill an exact constraint. Note that this is only possible since the graph is acyclic (otherwise the algorithm can output non-simple paths). More formally the update rule for the Bellman-Ford algorithm is the following:
$$ d(s, v, rm, rn) = \min_{u \in V}\{d(s, u, rm - \mathds{1}_{(u, v) \in R_m}, rn - \mathds{1}_{(u, v) \in R_n}) + \Bar{w}(u,v)|(u, v) \in E\}$$
where $G_C = (V,E)$, $\Bar{w}(u,v) = -w(e)$ for $e= (u,v)$, $R_m$ is the set of matching red edges in $G_C$ and $R_n$ is the set of non-matching red edges in $G_C$. We are interested in the value $d(s, t, |(C\backslash (t,s))\cap R_m|, |(C\backslash (t,s))\cap R_n|)$ where $|(C\backslash (t,s))\cap R_m|$ is the number of matching red edges in $C\backslash (t,s)$ and $|(C\backslash (t,s))\cap R_n|$ is the number of non-matching red edges in $C\backslash (t,s)$. 
The DP runs in polynomial time since the number of table entries is polynomial and allows us to find a simple path $path(C)$ (since the graph is acyclic) from $s$ to $t$ of minimum weight, i.e., $\Bar{w}(path(C)) \leq \Bar{w}(C\backslash (t,s))$ which implies $w(path(C)) \geq w(C\backslash (t,s))$, and with the same number of matching and non-matching red edges as $(C\backslash (t,s))$.

\proofsubparagraph*{Constructing the set $\mathcal{C}'$.} Finally for $C \in \mathcal{C}$, let $cycle(C) := path(C) \cup (t,s)$ with $path(C)$ computed using the above DP on the processed graph $G_C$. Let $\mathcal{C}' := \{cycle(C) | C \in \mathcal{C} \}$. Observe that $cycle(C)$ is a cycle with the same number of matching and non-matching red edges as $C$, $w(cycle(C)) \geq w(C)$ and all cycles in $\mathcal{C'}$ are $M$-alternating and disjoint. So $\mathcal{C'}$ fulfills all the required properties.

\proofsubparagraph*{Running Time.}
Observe that all the above steps can be run in $L^{O(L)}poly(n)$ time, so the total running time is of the same order.
\end{proof}




The above proposition is the key to proving \Cref{th:fptcirc} which gives an FPT algorithm for EM parameterized by the circumference of the graph. 
But first we need the following lemma which ensures that we can always make progress using a set of alternating cycles of size bounded by a function of their individual lengths.

\begin{lemma}\label{lem:circset}
Given a "Yes" instance of EM and a PM $M$, if $|R(M)|< k$ then there exists a set of disjoint $M$-alternating cycles $\mathcal{C}$ s.t. $|E(\mathcal{C})| \leq 2c^4$, where $c$ is circumference of the graph, and $|R(M)|<|R(M \Delta \mathcal{C})| \leq k$.
\end{lemma}

\begin{proof}
Let $M^*$ be a PM with $k$ red edges.
Note that for $\mathcal{C} = M \Delta M^*$ we get $|R(M)|<|R(M \Delta \mathcal{C})| = |R(M^*)| = k$. So at least one set of cycles fulfills the requirement on the number of red edges. Now let $\mathcal{C}$ be a minimum size set of cycles such that $|R(M)|<|R(M \Delta \mathcal{C})| \leq k$. We want to bound the number of cycles in $\mathcal{C}$.
We define new edge weights as follows: blue edges get weight
$0$, red edges in $M$ get weight $-1$ and red edges not in $M$ get weight $+1$. This way we have $|R(M \Delta \mathcal{C})| = |R(M)| + w(\mathcal{C})$.
Note that $w(\mathcal{C}) >0$ and $\mathcal{C}$ cannot contain any subset of cycles of total weight $0$ (this includes single cycles not containing red edges) since removing such a subset of cycles from $\mathcal{C}$ does not affect the total number of red edges in $\mathcal{C}$ and contradicts its minimality. Also note that every cycle $C \in \mathcal{C}$ contains at most $c$ edges (since $c$ is the circumference), so $|w(C)| \leq c$. This also implies that $w(\mathcal{C}) \leq c$ for otherwise we can remove any positive cycle from $\mathcal{C}$ while still having $|R(M)|<|R(M \Delta \mathcal{C})| \leq k$ thus contradicting its minimality. 

Let $\mathcal{C}^+$ be the set of positive cycles in $\mathcal{C}$ and $\mathcal{C}^-$ the set of negative cycles in $\mathcal{C}$.
Suppose $|\mathcal{C}^+| > c^3$. Then $|\mathcal{C}^-| > c^2-c$ since $w(\mathcal{C}) >0$. Observe that $\mathcal{C}^+$ must contain at least $c$ cycles of the same weight $w_1$ with $w_1 \leq c$ and $\mathcal{C}^-$ must contain at least $c$ cycles of the same weight $-w_2$ with $-w_2 \geq -c$. But then the set of cycles consisting of $w_1$ cycles of weight $-w_2$ and $w_2$ cycles of weight $w_1$ has total weight $0$, a contradiction. This means that $|\mathcal{C}^+| \leq c^3$. Similarly we get $|\mathcal{C}^-| \leq c^3$. So we have $|\mathcal{C}| \leq 2c^3$. Since each cycle in $\mathcal{C}$ has length at most $c$ edges we get $|E(\mathcal{C})| \leq 2c^4$. \end{proof}

\begin{proof}[Proof of \Cref{th:fptcirc}]
Let $M$ be a PM containing a minimum number of red edges (should be at most $k$). Note that $M$ can be computed in polynomial time by simply using a maximum weight perfect matching algorithm with $-1$ weights assigned to red edges and $0$ weights assigned to blue edges.
From \Cref{lem:circset} we know that there exists a set of disjoint $M$-alternating cycles $\mathcal{C}$ s.t. $|E(\mathcal{C})| \leq 2c^3$ and $|R(M)|<|R(M \Delta \mathcal{C})| \leq k$. Let $M' := M \Delta \mathcal{C}$. Now by using \Cref{prop:smallsetcycles} we can find a PM $M''$ with $|R(M'')| = |R(M')|$ (here we do not need to assign any weights to edges so the weight function used to apply the proposition can simply be uniform) so $|R(M)|<|R(M'')| \leq k$. We can repeat the procedure (applying \Cref{lem:circset} on $M''$) until we get a PM with exactly $k$ red edges. We need at most $k$ repetitions, each running in time $f(L)poly(n) = L^{O(L)}poly(n)$ for $L= O(c^3)$, i.e., we get an FPT algorithm parameterized by $c$.
\end{proof}

\Cref{th:fptcirc} illustrates the use the \cref{prop:smallsetcycles} to develop FPT algorithms for Exact Matching. However, the circumference of the graph can in general be quite large. 
We believe that \cref{prop:smallsetcycles} can be applied to get other more interesting FPT algorithms for EM and related problems. 
In \Cref{sec:TkPMFPT} we show one such application.

\section{Top-k Perfect Matching}\label{sec:TkPM}

In this section we study TkPM which, as we show later, is polynomial time equivalent to EM, making it another problem that can be used to test the hypothesis $\textbf{P} = \textbf{RP}$, but with the advantage of being an optimization problem.

\subsection{Minimum Weight Variant}\label{section-minW}

First, we start by introducing a variant of TkPM in which we are looking for a PM minimizing (instead of maximizing) the top-$k$ weight. This objective function has been studied in the context of other problems such as $k$-clustering and load balancing \cite{chakrabarty2019approximation} but to our knowledge, no prior work considered it in the context of matching problems.
\noindent\vspace{5pt}\begin{boxedminipage}{\textwidth}
\textsc{Minimum Weight Top-$k$ Perfect Matching (minTkPM)}

\textbf{Input:} A weighted graph $G$ and integer $k$.

\textbf{Task:} Find a perfect matching in $G$ minimizing the top-$k$ weight function.

\end{boxedminipage}
We show however, that by simply applying a threshold to the weights of the edges, we are able to reduce this problem to minimum weight perfect matching (minWPM), i.e., it is in \textbf{P}. 
The proof crucially relies on the idea of thresholding the weights which will also be useful for the approximation algorithms in the next sections.
\begin{definition}
Given a weighted graph $G$ with weights $w$, the thresholded weights $w_t$ for a threshold $t$ are defined as follows: for an edge $e$, $w_t(e)= \max{(w(e)-t,0)}$.
\end{definition}

\begin{theorem}\label{th:minTkPM}
$minTkPM \equiv_p minWPM$. 
\end{theorem}

\begin{proof}
$minWPM \leq_p minTkPM$ is trivial by setting $k=n/2$ so we need to prove $minTkPM \leq_p minWPM$. Given an instance of minTkPM, let $M^*$ be an optimal PM. Let $e_{k}$ be the $k$th edge from $M^*$ in the edge ordering.
The algorithm starts by guessing $e_k$ (i.e., running for all possibilities of $e_k$ and outputting the matching of smallest top-$k$ value among all solutions) and setting $t:=w(e_k)$.
We have $w_t(M^*) = w_t^k(M^*) = w^k(M^*) - k t$ since the $k$ values above $t$ are reduced by $t$, and the rest is set to $0$. Now let $M$ be a minimum weight perfect matching in the thresholded graph. Then we have $w_t^k(M)  \leq w_t(M) \leq w_t(M^*) \leq w(M^*) - k t$. After removing the threshold, each of the top-$k$ values can only increase by at most $t$, so we get $w^k(M)  \leq w(M^*)$, i.e., we get an optimal solution.
\end{proof}

This creates an interesting division between the minimization and maximization of the top-$k$ values, in the context of a perfect matching problem. On the one hand we have a problem that is polynomially equivalent to the general weighted matching problem (known to be in \textbf{P}), and on the other hand we get a problem that is polynomially equivalent to EM (as we show next) whose complexity remains unknown.

\subsection{Reducing Top-k Perfect Matching to Exact Matching}\label{sec:reduction}

To help reduce TkPM to EM, we introduce an intermediary problem called maximum weight EM in which we are given an instance of EM as well as edge weights (of polynomial size) and the goal is to find a PM with exactly $k$ red edges having maximum weight among all such PMs.

\vspace{5pt}
\noindent\vspace{5pt}\begin{boxedminipage}{\textwidth}
\textsc{Maximum Weight Exact Matching (MWEM)}

\textbf{Input:} An edge-weighted and edge-colored (with red/blue colors) graph $G$ and integer $k$.

\textbf{Task:} Find a perfect matching $M$ in $G$ with exactly $k$ red edges and having maximum weight among all such matchings.

\end{boxedminipage}
We show that this new variant can be reduced to EWPM (with polynomial weights) which in turn can be reduced to EM. This shows that MWEM is in $\textbf{RP}$.

\begin{lemma}\label{th:reductions}\hyperref[th:reductions:proof]{$(\star)$}
$MWEM \leq_p EWPM \leq_p EM$ for polynomially bounded weights. The reductions also work for bipartite input graphs and for minor closed graph classes.
\end{lemma}

\begin{proof}[Proof of \Cref{th:reductions}]\label{th:reductions:proof}
MWEM $\leq_p$ EWPM:
Given an instance of MWEM, let $w_{max}$ be the maximum edge weight value plus 1 and $M^*$ an optimal solution. Define new weights $w'$ such that blue edges keep the same weight $w'(e) := w(e)$ while red edges get weight $w'(e) := w(e) + nw_{max}$. Note that $w'(M^*) = w(M^*) + knw_{max}$. Observe that any PM $M$ of weight $(k+0.5)nw_{max} < w'(M) < (k+0.5)nw_{max}$ must contain exactly $k$ red edges since $ |w(M)| < 0.5nw_{max}$. 
Now we run a search algorithm for EWPM on the graph with weights $w'$ and weight parameter $W = (k+1)nw_{max}$. If the algorithm fails, we decrease $W$ by $1$ and repeat. Observe that the algorithm only succeeds when $W = w(M^*)$ and outputs an optimal solution for MWEM. 

EWPM $\leq_p$ EM (from \cite{gurjar2012planarizing}):
Given an instance of EWPM with parameter $W$,
replace every edge $e$ by a path of length $2w(e)-1$ with alternating red and blue colors starting and ending with red. Observe that every PM $M$ in the original graph corresponds to a PM $M'$ in the new graph such that if $e \in M$ then $M'$ contains the red edges of the path corresponding to $e$ (of which there are $w(e)$ many) and if $e \notin M$ then $M'$ contains only blue edges from the path corresponding to $e$. So the number of red edges in $M'$ would be equal to $w(M)$, which means that running an EM algorithm with $k= W$ on the new graph decides the EWPM instance.

For all reductions, observe that a bipartite input graph is transformed into a bipartite graph and an input graph from a minor closed family is transformed into a graph from the same family.
\end{proof}

Note that even though MWEM is an optimization problem, any approximation for it requires solving EM.
So our focus will instead be on TkPM which we reduce to EM when the input weights are polynomially bounded in the input size.

\begin{proof}[Proof of \Cref{th:TkPMtoEM}]\label{th:TkPMtoEM:proof}
We have $MWEM \leq_p EM$ from \Cref{th:reductions}, so we need to show that $TkPM \leq_p MWEM$.
Given an instance of TkPM, let $M^*$ be an optimal solution. Let $e_{k}$ be the $k$th edge from $M^*$ in the edge ordering.
The algorithm starts by guessing $e_k$ (i.e., running for all possibilities of $e_k$ and outputting the matching of highest top-$k$ value among all solutions) and setting the weights of all edges after $e_k$ in the ordering to 0 and coloring them blue, while the rest of the edges are colored red.
Note that only red edges can have non-zero weights and that $M^*$ has exactly $k$ red edges. Let $M$ be the output of an algorithm for MWEM on the resulting graph. By optimality of $M$, we have that $w(M) \geq w(M^*)$, and since they both contain at most $k$ non-zero weight edges we get $w^k(M) \geq w^k(M^*)$ so $M$ is an optimal solution for TkPM.
Since we only modify the weights of the edges, the reduction preserves the graph class.
\end{proof}

The above lemma, in combination with the result of \cite{elmaalouly2022exacttopkequiv}, implies the following theorem.

\begin{theorem}\label{th:EMequivTkPM}
$TkPM \equiv_p EM$ for polynomially bounded weights. The equivalence also holds for bipartite input graphs and for minor closed graph classes.
\end{theorem}
Note that it is still open whether MWEM and TkPM with exponential weights are reducible to EM or if they are \textbf{NP}-hard.




\subsection{Approximation Algorithms for Top-k Perfect Matching}\label{section-colored}

Note that the reduction to EM in \Cref{th:reductions} does not preserve any approximation factor since it changes the weights of the edges. So we cannot use it in combination with \Cref{th:EMapprox} to get an approximation algorithm for TkPM. We will, however, use \Cref{prop:constrainedCycle} to get a better approximation for TkPM as we will see later. 
First we show that by simply applying a specific threshold to the weights of the graph, any maximum weight perfect matching (maxWPM) algorithm can output a 0.5-approximation for TkPM.

\begin{lemma}%
\label{lem:goodthreshold}
Given an instance of TkPM, let $M^*$ be an optimal solution. There exists a threshold $t$ such that for any maximum weight perfect matching $M$ in the thresholded graph, we have $w^k(M) \geq 0.5\cdot w^k(M^*)$ (in the original graph).
\end{lemma}

\begin{proof}

Let $t=\frac{w^k(M^*)}{2k}$. Then we have $w_t(M^*) \geq w^k(M^*) - k\cdot \frac{w^k(M^*)}{2k} = 0.5 w^k(M^*)$. And since $M$ is a maximum weight perfect matching, we have $w_t(M) \geq w_t(M^*) \geq 0.5 w^k(M^*)$. Let $k'$ be the number of edges $e \in M$ with $w_t(e) > 0$. Now we have two cases. First, if $k' \leq k$ then we have $w^k(M) \geq w_t^k(M) \geq 0.5 w^k(M^*)$. Otherwise the output matching contains at least $k$ edge of weight more than $\frac{w^k(M^*)}{2k}$, so the total weight is $w^k(M) \geq k\cdot \frac{w^k(M^*)}{2k} = 0.5w^k(M^*)$.

 \end{proof}
 
The above lemma guarantees the existence of a threshold that will lead to a $0.5$-approximation using any maximum weight PM algorithm. We may not know the exact threshold, but if the weights are polynomial we can simply try all possibilities. Otherwise we can find a good threshold using binary search (see algorithm \ref{alg:05approx} in the appendix).  

\begin{algorithm}[hbt!]
\caption{TkPM 0.5-approximation Algorithm}\label{alg:05approx}
\KwIn{An instance of TkPM.}
\KwOut{PM $M$ with $w^k(M) \geq 0.5 w(M^*)$ where $M^*$ is an optimal solution.}

$t_1 \gets 0$, $t_2 \gets w_{max}$ \Comment*[r]{where $w_{max}$ is the maximum weight in the graph}

$M \gets \textsc{MaximumWeightPerfectMatching}(G,w_{t_1})$\;
\eIf{$M$ contains at most $k$ edges $e$ with $w(e) > 0$}{\KwRet{$M$}\;}{
    $M_1 \gets \textsc{MaximumWeightPerfectMatching}(G,w_{t_1})$\;
    $M_2 \gets \textsc{MaximumWeightPerfectMatching}(G,w_{t_2})$\;
    \While{$t_2-t_1 \geq 1/(k^2)$}{
        $t \gets (t_1+t_2)/2$ \;
        $M \gets \textsc{MaximumWeightPerfectMatching}(G,w_{t})$\;
        \eIf{$M$ contains more than $k$ edges $e$ with $w_t(e) > 0$}
        {$t_1 \gets t$;
        $M_1 \gets M$\;}{$t_2 \gets t$;
        $M_2 \gets M$\;}
        }
    $M \gets \textsc{BestOf}(M_1,M_2)$\;
    \KwRet{$M$}\;
}

\vspace{5pt}

\SetKwProg{Fn}{Procedure}{:}{end}
\Fn{\textsc{BestOf($M_1,M_2$)}{}}{
    \eIf{$w^k(M_1) \geq w^k(M_2)$}{\KwRet{$M_1$}\;}{\KwRet{$M_2$\;}}
}

\end{algorithm}

\begin{proof}[Proof of \Cref{th:05approx}]\label{th:05approx:proof}
Given an instance of TkPM, let $t_1 = 0$ and $t_2 = w_{max}$ where $w_{max}$ is the maximum weight in the graph. 
Let $M_{t_2}$ be the maximum weight perfect matching for threshold $t_2$. Note that all edges in the graph have weight 0, so $M_{t_2}$ has less than $k$ edges with non-zero weights.
Let $M_{t_1}$ be the maximum weight perfect matching for threshold $t_1$. Observe that if $M_{t_1}$ has at most $k$ edges with non-zero weights, then $M_{t_1}$ is also optimal with respect to the top-$k$ objective. So in this case the algorithm can simply output $M_{t_1}$. So we assume that $M_{t_1}$ has more than $k$ edges with non-zero weights.

Now using binary search, we can find a threshold $t$ for which we get a PM $M$ with at most $k$ edges $e$ with $w_{t}(e) >0$ and for $t' = t-1/(2k)$ \footnote{To keep the weights integral we can multiply all weights by $2k$.} we get a PM $M'$ containing more than $k$ edges $e$ with $w_{t'}(e) >0$ (see Algorithm \ref{alg:05approx}). The algorithm outputs the best of $M$ and $M'$ in terms of top-$k$ weight. Observe that $w^k(M) \geq w_t^k(M) = w_t(M) \geq w_t(M^*) \geq w(M^*) - kt$ and $w^k(M') \geq kt' = kt - 0.5$. Now if $kt \leq 0.5 w^k(M^*)$ then $w(M) \geq 0.5 w^k(M^*)$, otherwise $w^k(M') \geq kt - 0.5 > 0.5w^k(M^*) -0.5$ so $ w^k(M') \geq 0.5  w^k(M^*)$.
Note that the binary search takes at most $k \log_2(w_{max})$ steps, so the total running time is polynomial.
\end{proof}

In order to get a better approximation factor, we rely on \Cref{prop:constrainedCycle} which allows us to limit the change in the number of edges with weight above threshold.

\begin{proof}[Proof of \Cref{th:08approx}]
We start with a high level intuition on how the algorithm works and why it gives a better approximation. 

The core idea of the algorithm is the following: instead of recomputing the maximum weight perfect matching every time we change the threshold (as is done in the previous algorithm), we keep track of one perfect matching $M$ which we incrementally improve using alternating cycles that increase its weight. We also make sure that the cycles do not add too many edges of weight above the threshold. This way the top-$k$ weight of $M$ stays closer to its total weight. 
To find such cycles, we rely on the algorithm of \Cref{prop:constrainedCycle}, which allows us to find positive alternating cycles that do not add too many positive weight edges (at most $k$). But first we set the edge weights of edges in $M$ to negative (i.e., multiply them by $-1$). This way the weight of an alternating cycle indicates the total weight change we get when taking its symmetric difference with $M$ to get a new perfect matching.
This means that, whenever possible, we can increase the total weight of $M$ while keeping its top-$k$ weight close to its total weight (considering the thresholded weights) since the number of positive weight edges above the threshold is limited.

To see why this is helpful, consider the two cases in the proof of \Cref{lem:goodthreshold}: $k' \leq k$ and $k' \geq k$ (remember that $k'$ is the number of edges in $M$ with weight strictly above the threshold).

In the case $k' \leq k$ (let $k_1 := k'$), we know that the top-$k$ weight is the same as the total weight (for the thresholded weights), which means that we do not lose anything when considering only the top-$k$ weight (i.e., $w^k_t(M) = w_t(M) \geq w_t(M^*) = w^k_t(M^*)$). However, we might lose some value because of the threshold. This is because when we go back to the original weights, $M^*$ regains up to $k \cdot t$ in value ($k$ times the threshold, since all its top-$k$ edges might have value above the threshold) whereas $M$ might only regain $k_1 \cdot t$ (since all other edges could have original weight close to zero). So in the case $k_1 << k$, we lose almost all the value from the threshold. 

On the other hand, in case $k' \geq k$ (let $k_2 := k'$), $M$ will also regain $k \cdot t$ in top-$k$ weight when we add back the threshold. However, the top-$k$ weight of $M$ can be far from its total weight since many edges can be contributing to the total weight. This is mainly a problem when $k_2 >> k$. If $k'$ is close to $k$, however, this loss is not so big (at most a fraction $(k_2-k)/k$ of the total since the $k$ highest weights still count).

To get a worst case approximation factor of $0.5$, it must be the case that both $k_1 << k$ and $k_2 >> k$. Note, however, that the procedure detailed above (relying on \Cref{prop:constrainedCycle}) allows us to bound the difference between $k_1$ and $k_2$ by $k$ (i.e,. $k_2 - k_1 \leq k$). This way, the algorithm manages to guarantee a better approximation factor.

We are now ready to describe the full algorithm.
We will first show how to transform exponential weights into polynomially bounded ones while loosing at most a factor of $1-1/poly(n)$ in the approximation. We then provide a $0.8-1/poly(n)$-approximation algorithm for TkPM with polynomially bounded weights, which proves the theorem.

\proofsubparagraph*{Dealing with Exponential Weights.} For exponential size weights, we first scale and round them to make them bounded by a polynomial function of the input size. We start by deleting all edges that cannot be part of any perfect matching (this can simply be done by checking for every edge whether we could remove it along with its endpoints from the graph and still be able to get a perfect matching on the rest of the graph). Let $W$ be the highest edge weight in the remaining graph. 
Observe that for $k \geq 1$ an optimal solution $M^*$ to the top-$k$ perfect matching problem must have $w^k(M^*) \geq W$ (since the perfect matching containing the edge of weight $W$ is a valid solution).
Now all weight encodings have at most $\log_2(W)$ non-zero bits. Let $f(n) = n\cdot poly(n)$ for any desired polynomial. If $W \leq f(n)$ then all weights are polynomial. Otherwise we re-encode the weights of all edges by only considering their $(\log_2(W)-\log_2(f(n)))$-th to $(\log_2(W))$-th bits (counting from the least significant bit) and dropping all others. We call these weights $w'$. So all weights are now encoded with at most $\log_2(f(n)) + 1$ bits, i.e., are bounded by a polynomial function of the input size. 
Now let $M'$ be a $0.8-1/poly(n)$-approximation for TkPM on the graph with weights $w'$. Observe that for any edge $e$, $|w(e) - w'(e)\cdot\frac{W}{f(n)}| \leq \frac{W}{f(n)}$ (the rounding error). 
Since a perfect matching contains $n/2$ edges we get
$$w^k(M') \geq w'^k(M')\cdot\frac{W}{f(n)} - \frac{nW}{2f(n)} \geq w'^k(M')\cdot\frac{W}{f(n)}(1 - \frac{n}{f(n)}).$$
The last inequality resulting from the fact that the optimal solution has weight at least $W \geq f(n)$ so $w'^k(M') \geq f(n)/2$. Now since $M'$ is a $0.8-1/poly(n)$-approximation we get
$$w^k(M') \geq (0.8-1/poly(n)) \cdot w'^k(M^*)\cdot\frac{W}{f(n)}(1 - \frac{n}{f(n)})  \geq (0.8-2/poly(n)) \cdot w'^k(M^*)\cdot\frac{W}{f(n)}.$$ 
Going back to the original weights, we get
$$w^k(M')  \geq (0.8-2/poly(n)) \cdot (w^k(M^*) - \frac{nW}{2f(n)}).$$ 
Finally, using $w^k(M^*) \geq W \geq f(n)$ we get $$w^k(M') \geq  (0.8-3/poly(n)) w^k(M^*).  $$


\proofsubparagraph*{Approximation algorithm for polynomial weights.}
We start with a preprocessing of the edge weights.
Given an instance of TkPM, let $M^*$ be an optimal solution. Let $e_{k}$ be the $k$th edge from $M^*$ in the edge ordering.
The algorithm starts by guessing $e_k$ (i.e., running for all possibilities of $e_k$ and outputting the matching of highest top-$k$ value among all solutions) and setting the weights of all edges after $e_k$ in the ordering to 0.
This means that some solution $M^*$ will contain at most $k$ edges of strictly positive weight. Note that this does not modify the top-$k$ weight of $M^*$ and can only decrease the top-$k$ weight of any matching, i.e., if we find a $0.8-1/poly(n)$-approximation on the new weights, it is also a $0.8-1/poly(n)$-approximation on the original weights.
Also note that now we have $w^k(M^*) = w(M^*)$.
For ease of notation, let $\epsilon = 1/poly(n)$ for the desired polynomial function in the approximation factor.
The algorithm will have two phases.
In the first phase, we use a slight modification of algorithm \ref{alg:05approx} (we call it \textsc{ModifiedTkpmApprox} in algorithm \ref{alg:08approx}) where instead of returning $M$, it returns $M_1$, $M_2$ and $t_2$. We label them $M_0'$, $M_0$ and $t_0$ respectively. This way we are guaranteed that $M_0$ has at most $k$ edges with $w_{t_0}(e) >0$ and $M_0'$ has at least $k$ edges with $w_{t_0-\epsilon /k}(e) >0$, i.e., we have one matching with less then $k$ edges of weight strictly more than $t_0$ and one with more than $k$ edges of weight strictly more than $t_0-\epsilon /k$. Algorithm \ref{alg:05approx} also guarantees that both matchings have maximum total weight with respect to their thresholds.

The second phase only works for bipartite graphs since it will rely on the algorithm of \Cref{prop:constrainedCycle} (see \textsc{ImproveUsingBoundedCycles} in algorithm \ref{alg:08approx}) to increase the weight of the matching instead of recomputing the maximum weight perfect matching from scratch whenever we decrease the threshold (as is done in algorithm \ref{alg:05approx}). This way we again search for a threshold $t_1$ such that \textsc{ImproveUsingBoundedCycles} fails to get a PM containing more than $k$ edges $e$ with $w_{t_1}(e) >0$ and instead stops at a perfect matching $M_{t_1}$ containing at most $k$ edges $e$ with $w_{t_1}(e) >0$, but for threshold $t_2 = t_1 - \epsilon /k$ \textsc{ImproveUsingBoundedCycles} outputs two perfect matchings $M_1$ and $M_2$ such that $M_2$ has at most $k$ edges $e$ with $w_{t_2}(e) >0$ and $M_1$ has more than $k$ edges $e$ with $w_{t_2}(e) >0$.

The full algorithm is given in algorithm \ref{alg:08approx}. Note that for a graph $G := (A \cup B, E, w)$ and a perfect matching $M$, we define the directed graph $G_M := (A \cup B, E, w')$ where every edges in $M$ is oriented from $A$ to $B$ and has weight $w'(e) = -w(e)$ and every edge not in $M$ is oriented from $B$ to $A$ and has weight $w'(e) = w(e)$. This means that for $M' := M \Delta C$ where $C$ is a directed cycle (which implies that it is $M$-alternating), $M'$ is a PM with $w(M') = w(M) + w'(C)$.
\begin{algorithm}[hbt!]
\caption{TkPM 0.8-approximation Algorithm}\label{alg:08approx}
\KwIn{An instance of TkPM and $\epsilon := 1/poly(n)$ for some polynomial function $poly(n)$.}
\KwOut{PM $M$ with $w^k(M) \geq (0.8-\epsilon) w(M^*)$ where $M^*$ is an optimal solution.}

\For{$e_k \in E(G)$}{
    \For{$e \in E(G)$}{\If{$e > e_k$\Comment*[r]{according to the edge ordering of the graph}} 
        {
        $w(e) \gets 0$\;}{}}

    $M_0', M_0, t_0 \gets \textsc{ModifiedTkpmApprox}$\;
    $M_1,M_2 \leftarrow M_0$\;
    $Success \gets False$\;
    
    \While{Success == False and  $t - \epsilon /k > 0$}{
        $t \gets t - \epsilon /k$\;
        $M_1, M_2, Success \gets $ \textsc{ImproveUsingBoundedCycles}($M_1,t$)\;
        \eIf{Success}{
            $t_2 \gets t$\;
            $t_1 \gets t+\epsilon /k$\;} 
            {$M_0 \gets M_1$\;}
        }

    $M \gets \textsc{BestOf}(M,M_0,M_0',M_1,M_2)$\;
}
\KwRet{$M$}\;


\vspace{5pt}

    \SetKwProg{Fn}{Procedure}{:}{end}
    \Fn{\textsc{ImproveUsingBoundedCycles($M,t$)}{}}{
    \eIf{$M_1$ contains more than $k$ edges $e$ with $w_t >0$}{
            \textbf{return} $ M_1, M_1, True$\;
        }{
    $M_1, M_2 \leftarrow M$\;
        \While{$M_1$ contains at most $k$ edges $e$ with $w_t(e) >0$ and there exists a positive cycle in $G_{M_1}$, for threshold $t$, containing at most $k$ edges $e$ with $w_t(e) >0$ and $e \notin M_1$}{
            Use \Cref{prop:constrainedCycle} to find a cycle $C$ with the above properties\;
            $M_2 \leftarrow M_1$\;
            $M_1 \leftarrow M_2 \Delta C$\;
            }
        \eIf{$M_1$ contains less than $k$ edges $e$ with $w_t(e) >0$}{
            \textbf{return} $ M_1, M_1, False$\;
        }{\textbf{return} $ M_1, M_2, True$\;}
        }
}

\end{algorithm}

\proofsubparagraph*{Proof of Correctness.}
The algorithm outputs the best (in terms of top-$k$ weight) among $M_0$, $M_0'$, $M_1$ and $M_2$. So we only need to show that
$$\max{(w^k(M_0),w^k(M_0'),w^k(M_1),w^k(M_2))} \geq 0.8 w(M^*)- \epsilon.$$
Note that we only need to prove correctness for the correct guess of $e_k$. Observe that for a perfect matching $M$, the symmetric difference between $M$ and $M^*$ (which consists of disjoint $M$-alternating cycles) forms a set of directed cycles in $G_M$ each containing no more than $k$ edges of strictly positive weight. This is because edges in $M$ have their weight sign flipped (so they have negative weight) and $M^*$ contains at most $k$ edges of strictly positive weight. Also note that if for a threshold $t$ we have $w_t(M) < w_{t}(M^*)$, then $M \Delta M^*$ contains at least one strictly positive weight cycle (with respect to $w_t$) since $w_{t}(M^*) = w_{t}(M) + w_t(M \Delta M^*)$. This means that the conditions of \Cref{prop:constrainedCycle} are met and \textsc{ImproveUsingBoundedCycles} cannot fail in this case. 

Suppose \textsc{ImproveUsingBoundedCycles} always fails, even for threshold $t<\epsilon /k$. This means the the conditions of \Cref{prop:constrainedCycle} are not met so 
$w_t(M_1) \geq w_t(M^*)$ (as argued above) and since $M_1$ contains less than $k$ strictly positive weight edges we have 
$$w_t^k(M_1) = w_t(M_1) \geq w_t(M^*) \geq (0.8- \epsilon)w(M^*).$$

Now suppose \textsc{ImproveUsingBoundedCycles} succeeds at least once. Suppose the first time this happens is at threshold $t_2$.
Since \textsc{ImproveUsingBoundedCycles} fails for threshold $t_1 = t_2 + \epsilon/k$, the outputted matching $M_{t_1}$ should have less than $k$ edges $e$ with $w_{t_1}(e) >0$ and $w_{t_1}(M_{t_1}) \geq w_{t_1}(M^*)$.
Note that \textsc{ImproveUsingBoundedCycles} either succeeds in the first "If" condition, i.e., $M_1$ contains more than $k$ edges $e$ with $w_{t}(e) >0$ (where in this case $t=t_2$) or later after the "While" loop. In the former case we also know that $M_1$ (which is equal to $M_{t_1}$) had less than $k$ edges $e$ with $w_{t_1}(e) >0$ so we have 
$$w^k_{t_1}(M_1) \geq w_{t_1}(M_1) = w_{t_1}(M_{t_1}) \geq w_{t_1}(M^*).$$
Since $M_1$ contains more than $k$ edges $e$ with $w_{t_2}(e) >0$ we also have
$$w^k(M_1) \geq w^k_{t_2}(M_1) + k t_2 \geq w^k_{t_1}(M_1) + k t_2.$$
This implies 
$$w^k(M_1) \geq w^k_{t_1}(M_1) + k t_1 - \epsilon \geq w_{t_1}(M^*) + k t_1 - \epsilon \geq  w^k(M^*)- \epsilon \geq (0.8- \epsilon) w^k(M^*).$$

In the latter case (i.e., success after the "While" loop), let $k_1$ be the number of edges $e\in M_1$ with $w_{t_2}(e) >0$ and $k_2$ be the number of edges $e \in M_2$ with $w_{t_2}(e) >0$. Note that $k_1 \leq k$, $k_2 \geq k$ and $(k_2 - k_1) \leq k$ (due to the constraint on the cycle $C$).  
Since only strictly positive weight cycles were used to get to $M_2$ from $M_1$ and to $M_1$ from $M_{t_1}$ we get
$$w_{t_2}(M_2) \geq w_{t_2}(M_1) \geq w_{t_2}(M_{t_1}) \geq w_{t_1}(M_{t_1}) \geq w_{t_1}(M^*) \geq w_{t_2}(M^*) - \epsilon.$$
Now we have: 
$$w_{t_2}^k(M_1) = w_{t_2}(M_1) \geq w_{t_2}(M^*) - \epsilon \geq w(M^*) - k{t_2} - \epsilon$$ 
and 
$$w_{t_2}^k(M_2) \geq \frac{k}{k_2}w_{t_2}(M_2) \geq \frac{k}{k_2} w_{t_2}(M^*) - \epsilon \geq \frac{k}{k_2} (w(M^*) - k{t_2}) - \epsilon.$$
After removing the thresholds we get 
$$w^k(M_1) \geq w(M^*) - k{t_2} + k_1{t_2} - \epsilon \geq  w(M^*) - (k - k_1){t_2}- \epsilon$$ 
and
$$w^k(M_2) \geq \frac{k}{k_2}(w(M^*) - k{t_2}) - \epsilon + k{t_2} \geq w(M^*) - \frac{k_2-k}{k_2}(w(M^*) - k{t_2})- \epsilon.$$
%
Now suppose 
$$\max{(w^k(M_0),w^k(M_0'),w^k(M_1),w^k(M_2))} < 0.8 w(M^*)- \epsilon.$$
Observe that the threshold $t_0$ after the first phase must be at most $\frac{0.8w(M^*)}{k} $ (otherwise $w(M_0') \geq 0.8 w(M^*)$). 
Now we have 
$$w(M^*)- (k - k_1){t_2} - \epsilon \leq w^k(M_1) < 0.8 w(M^*)- \epsilon.$$ 
Combined with $k_1 \geq k_2-k$ we also get
$$t_2 \geq  \frac{0.2 w(M^*)}{2k - k_2}$$ 
and
$$w^k(M_2) \geq w(M^*)(1 - \frac{k_2-k}{k_2}(w(M^*) - k{t_2}) )- \epsilon \geq w(M^*)(1 - \frac{k_2-k}{k_2}(1-\frac{0.2k}{2k-k_2}) ) - \epsilon.$$
Now given that ${t_2} \leq t_0 \leq \frac{0.8w(M^*)}{k}$, we have $k_2 \leq 1.8 k$. By taking $k_2 = xk$, we get 
$$w^k(M_2)+ \epsilon \geq w(M^*)(1 -\frac{(x-1)(1.8-x)}{x(2-x)}) \geq 0.8 w(M^*)$$
where the last inequality comes from evaluating the function for all possible $x \in [1,1.8]$. 

So we get a contradiction, which means that we must have 
$$\max{(w^k(M_0),w^k(M_0'),w^k(M_1),w^k(M_2))} \geq 0.8 w(M^*)- \epsilon.$$

\proofsubparagraph*{Running Time.}
First we look at \textsc{ImproveUsingBoundedCycles}. Every time the "While" loop is repeated, the weight of the matching $M_1$ strictly increases, so it can only be run a polynomial number of times (since the weights are polynomial). Each run of the "While" loop is also polynomial in time since the algorithm of \Cref{prop:constrainedCycle} runs in polynomial time.
Now we look at the "While" loop in the main part of Algorithm \ref{alg:08approx}. Every time it is repeated, the threshold $t$ drops by $\epsilon /k$, so it can only be repeated $k\cdot W/\epsilon$ times, i.e., polynomial if the weights are polynomial and $\epsilon$ is inverse polynomial. 
So all steps in Algorithm \ref{alg:08approx} run in polynomial time and are repeated a polynomial number of times, i.e., the total running time is polynomial.
\end{proof}





\subsection{Parametrized Complexity of TkPM} \label{sec:TkPMFPT}

In this section we show that TkPM can be solved in Fixed Parameter Tractable (FPT) time parameterized by $k$ and $\alpha$, the independence number of the graph (i.e., the size of the largest independent set). The algorithm mainly uses \Cref{prop:smallsetcycles}. To prove its correctness we will rely 
on what \cite{elmaalouly2022exact} defines as skip, which allows us to shorten alternating cycles using the bound on the independence number. This way we manage to bound the total number of edges, in a symmetric difference with some optimal solution, by a function of $k$ and $\alpha$.
We start by adapting the following definition and lemma from \cite{elmaalouly2022exact}.

\begin{definition}\label{def:skip} (adapted from \cite{elmaalouly2022exact})
Let $C$ be a an $M$-alternating cycle. A skip $S$ is a set of 2 non-matching edges $e_1 := (v_1,v_2)$ and $e_2 := (v_1',v_2')$ with $e_1, e_2 \notin C$ and $v_1,v_1',v_2,v_2' \in C$ s.t. $C' = e_1 \cup e_2 \cup C \setminus (C[v_1,v_1'] \cup C[v_2,v_2'])$ is an $M$-alternating cycle and $|C| - |C'| > 0$. 
\end{definition} 

\begin{lemma}\label{lem:skipfrompaths}
(adapted from \cite{elmaalouly2022exact})
Let $P$ be an $M$-alternating path containing a set $\mathcal{P}$ of disjoint paths, each of length at least $3$ starting and ending at non-matching edges, and $|\mathcal{P}| \geq 4^\alpha$. Then $P$ contains a skip.
\end{lemma}
This allows us to prove the following.

\begin{proposition}\label{prop:cycleshortening}
Let $M$ be a PM in a weighted graph $G$ of independence number $\alpha$, with edge colors red and blue. Let $\mathcal{C}$ be a set of disjoint $M$-alternating cycles in $G$. Let $R \subseteq E(\mathcal{C})$ be the set of red edges in $\mathcal{C}$ and $k = |R|$. Then $G$ must contain a set of disjoint $M$-alternating cycles $\mathcal{C}'$ s.t.  $R \subseteq \mathcal{C}'$,  $|E(\mathcal{C})| \leq kf(\alpha)$ for $f(\alpha) = 4^{\alpha+1}$ and 
all edges in $E(\mathcal{C}') \backslash E(\mathcal{C})$ are non-matching edges.
\end{proposition}

\begin{proof}
We start with $\mathcal{C}' = \mathcal{C}$.
First observe that any cycle in $\mathcal{C}$ not containing any edge from $R$ can be removed from $\mathcal{C}'$. Next we show that any path $P \subseteq C \in \mathcal{C}'$ s.t. $|P| \geq f(\alpha)$ and $P\cap R = \emptyset $ can be shortened (i.e., replaced by a path of strictly smaller length) while keeping all the desired properties of the $\mathcal{C}'$. This is the result of applying \Cref{lem:skipfrompaths} to find a skip $S$ in $P$ and using it to shorten the cycle. To do so we need to show that $P$ contains a set $\mathcal{P}$ of disjoint paths, each of length at least $3$ starting and ending at non-matching edges, and $|\mathcal{P}| \geq 4^{\alpha+1}$. Observe that this is possible by simply splitting $P$ in paths of length $4$ and cutting off a matching edge from each of them. 
Note that using a skip can only add non-matching edges to a cycle. Now by repeatedly applying this shortening procedure as long as there exists a $P$ s.t. $|P| \geq f(\alpha)$ and $P\cap R = \emptyset $, we are left with a set of cycles $\mathcal{C}'$ containing no such path. Observe that in $\mathcal{C}'$ any two edges from $R$ can be separated by at most $f(\alpha)$ edges so $|E(\mathcal{C})| \leq kf(\alpha)$ as desired.
\end{proof}

\begin{lemma}\label{lem:smallSymDif}
Given an instance of TkPM and a PM $M$, there exists an optimal solution $M'$ s.t. $|E(M \Delta M')| \leq kf(\alpha)$ for $f(\alpha) = 4^{\alpha+1}$, where $\alpha$ is the independence number of the input graph.  
\end{lemma}

\begin{proof}
Let $M^*$ be an optimal solution. Let $e_{k}$ be the $k$th edge from $M^*$ in the strict ordering of the weights (where ties are broken arbitrarily). We color all edges of $G$ after $e_k$ in the ordering blue, and all other edges red. Note that $M^*$ has $k$ red edges. Observe that $\mathcal{C} = M \Delta M^*$ fulfills the requirements of \Cref{prop:cycleshortening} so there must a set of $M$-alternating cycles $\mathcal{C}'$ in G s.t. $R(\mathcal{C})\subseteq R(\mathcal{C}')$, $|E(\mathcal{C}')| \leq kf(\alpha)$, and any edge $e$ s.t. $e \in E(\mathcal{C}')$ and $e \notin E(\mathcal{C})$ is a non-matching edge. Let $M' = M \Delta \mathcal{C'}$.
\begin{claim*}
$R(M^*) \subseteq R(M')$.
\end{claim*} 
\begin{claimproof}
Let $e \in R(M^*)$. If $e \in M$ then $e \notin \mathcal{C}$ (since $\mathcal{C} = M \Delta M^*$). This implies that $e \notin \mathcal{C'}$ since since $e$ is a matching edge so $e \in M'$. 
Now if $e \notin M$ then $e \in \mathcal{C}$ and since $e$ is a red edge we get $e \in \mathcal{C}'$ which in turn implies $e \in M'$. 
 \end{claimproof} 
Now observe that $w^k(M') \geq w^k(M^*)$ since $M'$ contains all the edges of $M^*$ whose edge weights are counted in $ w^k(M^*)$ (i.e., the red edges). So $M'$ is an optimal solution and $|E(M \Delta M')| =|E(\mathcal{C}')| \leq kf(\alpha)$ as desired.
\end{proof}

\begin{proof}[Proof of \Cref{th:FPTkalpha}]\label{th:FPTkalpha:proof}
Given an instance of TkPM, we start by computing any perfect matching $M$. By \Cref{lem:smallSymDif}, we know that there exists an optimal solution $M'$ s.t. $|E(M \Delta M')| \leq kf(\alpha)$ for $f(\alpha) = 4^{\alpha+1}$. Let $e_{k}$ be the $k$th edge from $M'$ in the edge ordering.
The algorithm starts by guessing $e_k$ (i.e., running for all possibilities of $e_k$ and outputting the matching of highest top-$k$ value among all solutions) and setting the weights of all edges after $e_k$ in the ordering to 0 and coloring them blue, while the rest of the edges are colored red. Note that this does not modify the top-$k$ weight of $M'$ and can only decrease the top-$k$ weight of any other matching.
Now we can use the algorithm of \Cref{prop:smallsetcycles} to find a PM $M''$ with $w(M'') \geq w(M')$ and $|R(M'')| = |R(M')| = k$. Since only red edges have non-zero weight we get $w^k(M'') \geq w^k(M')$ so $M''$ is an optimal solution (since $M'$ is an optimal solution).

The running time is dominated by the algorithm of \Cref{prop:smallsetcycles} which runs in time $L^{O(L)}$ where $L = |E(M \Delta M')| \leq k4^{\alpha+1}$, and we get the desired running time.
\end{proof}


\subsubsection{Bipartite Case.}
In this section we prove \Cref{th:FPTkbeta}, the bipartite analogue of \Cref{th:FPTkalpha}.
Note that the existence of an FPT algorithm can be proven similarly to \Cref{th:FPTkalpha} using the concept biskip from \cite{elmaalouly2022exact} to replace the use of skips. \Cref{th:FPTkbeta}, however, gives an algorithm with a much better dependence on $\beta$ by instead relying on something we call a shortcut, which also allows for the shortening of cycles. 

We will rely on an orientation of the edges of the graph defined as follows. 
Given a bipartite graph $G$ with bipartition $(A,B)$ and a matching $M$, we transform $G$ into a directed graph $G_M$ by orienting every matching edge from $A$ to $B$ and every non-matching edge from $B$ to $A$.

\begin{definition}
Let $C$ be a directed $M$-alternating cycle. A shortcut $S$ is an edge $e := (v_1,v_2)$ with $e \notin C$ and $v_1,v_2 \in C$ s.t. $C' := C[v_2,v_1] \cup e $ is an $M$-alternating cycle and $|C'| < |C|$.
\end{definition}

\begin{lemma}\label{lem:shortcut}
Let $P := (v_1,v_2,...) $ be a directed path of length $|P| \geq 2\beta + 2$ contained in a directed cycle $C \subseteq G_M$. Then $P$ must contain a shortcut.
\end{lemma}

\begin{proof}
Let $V_A := P \cap A$ and $V_B:= P \cap B$. Note that $|V_A| = |V_B| = \beta +1$. Let $V_1$ be the first $\beta/2$ vertices in $V_B$ and $V_2$ the last $\beta/2$ vertices in $V_A$. Observe that $V_1 \cup V_2$ is a balanced set of size $\beta$, so there must be an edge $e = (v_i,v_{i'})$ connecting a vertex of $v_i \in V_1$ and a vertex of $v_{i'} \in V_2$ ($e$ must be directed from $V_1$ to $V_2$ since it corresponds to a non-matching edge). Observe that $i < i'$ so $C' := C[v_i,v_{i'}] \cup e $ is an $M$-alternating cycle and $|C'| < |C|$.
\end{proof}
This allows us to prove the following.

\begin{proposition}\label{prop:cycleshorteningBi}
Let $M$ be a PM in a weighted bipartite graph $G$ of bipartite independence number $\beta$, with edge colors red and blue. Let $\mathcal{C}$ be a set of disjoint $M$-alternating cycles in $G$. Let $R \subseteq E(\mathcal{C})$ be the set of red edges in $\mathcal{C}$ and $k = |R|$. Then $G$ must contain a set of disjoint $M$-alternating cycle $\mathcal{C}'$ s.t. $R \subseteq \mathcal{C}'$,  $|E(\mathcal{C})| \leq kf(\beta)$ for $f(\beta) = 2\beta + 2$ and all edges in $E(\mathcal{C}') \backslash E(\mathcal{C})$ are non-matching edges.
\end{proposition}

\begin{proof}
The proof is similar to that of \Cref{prop:cycleshortening} but uses \Cref{lem:shortcut} instead of \Cref{lem:skipfrompaths}.

We start with $\mathcal{C}' = \mathcal{C}$.
First observe that any cycle in $\mathcal{C}$ not containing any edge from $R$ can be removed from $\mathcal{C}'$. Next we show that any path $P \subseteq C \in \mathcal{C}'$ s.t. $|P| \geq f(\beta)$ and $P\cap R = \emptyset $ can be shortened (i.e., replaced by a path of strictly smaller length) while keeping all the desired properties of the $\mathcal{C}'$. This is the result of applying \Cref{lem:shortcut} to find a shortcut $S$ in $P$ and using it to shorten the cycle.
Note that using a shortcut can only add non-matching edges to a cycle. Now by repeatedly applying this shortening procedure as long as there exists $P \subseteq \mathcal{C}'$ s.t. $|P| \geq f(\beta)$ and $P\cap R = \emptyset $, we are left with a set of cycles $\mathcal{C}'$ containing no such path. Observe that in $\mathcal{C}'$ any two edges from $R$ can be separated by at most $f(\beta)$ edges so $|E(\mathcal{C})| \leq kf(\beta)$ as desired.
\end{proof}

\begin{lemma}\label{lem:smallSymDifBi}
Given a "Yes" instance of TkPM on a bipartite graph and a PM $M$, there exists an optimal solution $M'$ s.t. $|E(M \Delta M')| \leq kf(\beta)$ for $f(\beta) = 2\beta + 2$, where $\beta$ is the bipartite independence number of the input graph.  
\end{lemma}

\begin{proof}
The proof is similar to that of \Cref{lem:smallSymDif} but uses \Cref{prop:cycleshorteningBi} instead of \Cref{prop:cycleshortening}. 

Let $M^*$ be an optimal solution. Let $e_{k}$ be the $k$th edge from $M^*$ in the strict ordering of the weights (where ties are broken arbitrarily). We color all edges of $G$ after $e_k$ in the ordering blue, and all other edges red. Note that $M^*$ has $k$ red edges. Observe that $\mathcal{C} = M \Delta M^*$ fulfills the requirements of \Cref{prop:cycleshorteningBi} so there must a set of $M$-alternating cycles $\mathcal{C}'$ in G s.t. $R(\mathcal{C})\subseteq R(\mathcal{C}')$, $|E(\mathcal{C}')| \leq kf(\beta)$, and any edge $e$ s.t. $e \in E(\mathcal{C}')$ and $e \notin E(\mathcal{C})$ is a non-matching edge. Let $M' = M \Delta \mathcal{C'}$.
\begin{claim*}
$R(M^*) \subseteq R(M')$.
\end{claim*} 
\begin{claimproof}
Let $e \in R(M^*)$. If $e \in M$ then $e \notin \mathcal{C}$ (since $\mathcal{C} = M \Delta M^*$). This implies that $e \notin \mathcal{C'}$ since since $e$ is a matching edge so $e \in M'$. 
Now if $e \notin M$ then $e \in \mathcal{C}$ and since $e$ is a red edge we get $e \in \mathcal{C}'$ which in turn implies $e \in M'$. 
 \end{claimproof} 
Now observe that $w^k(M') \geq w^k(M^*)$ since $M'$ contains all the edges of $M^*$ whose edge weights are counted in $ w^k(M^*)$ (i.e., the red edges). So $M'$ is an optimal solution and $|E(M \Delta M')| =|E(\mathcal{C}')| \leq kf(\beta)$ as desired.
 \end{proof}

\begin{proof}[Proof of \Cref{th:FPTkbeta}]\label{th:FPTkbeta:proof}
The proof is similar to that of \Cref{th:FPTkalpha} but uses \Cref{lem:smallSymDifBi} instead of \Cref{lem:smallSymDif} and $\beta$ instead of $\alpha$.

Given an instance of TkPM, we start by computing any perfect matching $M$. By \Cref{lem:smallSymDifBi}, we know that there exists an optimal solution $M'$ s.t. $|E(M \Delta M')| \leq kf(\beta)$ for $f(\beta) = 2\beta + 2$. Let $e_{k}$ be the $k$th edge from $M'$ in the edge ordering.
The algorithm starts by guessing $e_k$ (i.e., running for all possibilities of $e_k$ and outputting the matching of highest top-$k$ value among all solutions) and setting the weights of all edges after $e_k$ in the ordering to 0 and coloring them blue, while the rest of the edges are colored red. Note that this does not modify the top-$k$ weight of $M'$ and can only decrease the top-$k$ weight of any other matching.
Now we can use the algorithm of \Cref{prop:smallsetcycles} to find a PM $M''$ with $w(M'') \geq w(M')$ and $|R(M'')| = |R(M')| = k$. Since only red edges have non-zero weight we get $w^k(M'') \geq w^k(M')$ so $M''$ is an optimal solution (since $M'$ is an optimal solution).

The running time is dominated by the algorithm of \Cref{prop:smallsetcycles} which runs in time $L^{O(L)}$ where $L = |E(M \Delta M')| \leq k4^{2\beta + 2}$, and we get the desired running time.
\end{proof}

\section{Conclusion and Open Problems}\label{sec:conc}

In the paper we study the Top-$k$ perfect matching problem which is shown to be polynomially equivalent to the Exact Matching problem. In the course of developing approximation algorithms for this problem we also initiate a new direction of study for the EM problem where the goal is to minimize the constraint violation while requiring the output to be a perfect matching.
We also continue the study of the parameterized complexity of EM that was initiated by \cite{elmaalouly2022exact}. To show the utility of these developments, we provide FPT algorithms for TkPM which rely on them. 

Our work leaves open many questions, we list a few. Starting with questions related to EM, can we reduce the constraint violation in \Cref{th:EMapprox} to less than $0.5k$? The aim would be to get $o(k)$, but so far no constant improvement is known. Also, can we design an FPT algorithm for EM parameterized by $k$ and $\alpha$? This would be an intermediate step towards an FPT algorithm parameterized only by $\alpha$ and resolving the open problem in \cite{elmaalouly2022exact}. For TkPM, the first open question is whether the problem is \textbf{NP}-hard. We know it is unlikely for the case of polynomial sized input weights since we can reduce it to EM, but the exponential size weights case is still open. Another interesting problem is to get an FPT (or even XP) algorithm for TkPM parameterized only by $\alpha$. Note that this is known for EM, but the reduction of TkPM to EM does not preserve the independence number of the graph. An FPT algorithm parameterized only by $k$, for either EM or TkPM would also be highly desirable. Finally an important step forward would be to improve the approximation algorithms for TkPM, with the goal of getting a polynomial time approximation scheme.





\bibliography{references}


\end{document}